\documentclass[reprint,twocolumn,aps,prl,nofootinbib,superscriptaddress]{revtex4-2}

\usepackage{amsmath}
\usepackage{amssymb}
\usepackage{mathrsfs}
\usepackage[dvipsnames]{xcolor}
\usepackage{graphicx}
\usepackage{hyperref}
\usepackage{times}

\newcommand{\eg}{e.g.,~}
\newcommand{\ie}{i.e.,~}

\hypersetup{colorlinks=true,        
linkcolor=blue,         
citecolor=cyan,         
}

\newcommand{\rom}[1]{\uppercase\expandafter{\romannumeral #1\relax}}

\begin{document}

\title{Impact of bulk viscosity on the post-merger gravitational-wave
  signal from merging neutron stars}

\author{Michail Chabanov}
\affiliation{Institut f\"ur Theoretische Physik, Goethe Universit\"at,
  Max-von-Laue-Str. 1, 60438 Frankfurt am Main, Germany}
\affiliation{Center for Computational Relativity and Gravitation \& School of Mathematical
Sciences, Rochester Institute of Technology, 85 Lomb Memorial Drive, Rochester,
New York 14623, USA}

\author{Luciano Rezzolla}
\affiliation{Institut f\"ur Theoretische Physik, Goethe Universit\"at,
  Max-von-Laue-Str. 1, 60438 Frankfurt am Main, Germany}
\affiliation{Frankfurt Institute for Advanced Studies,
  Ruth-Moufang-Str. 1, 60438 Frankfurt am Main, Germany}
\affiliation{School of Mathematics, Trinity College, Dublin 2, Ireland}

\date{\today}

\begin{abstract}
  In the violent post-merger of binary neutron-star mergers strong
  oscillations are present that impact the emitted gravitational-wave
  (GW) signal. The frequencies, temperatures and densities involved in
  these oscillations allow for violations of the chemical equilibrium
  promoted by weak-interactions, thus leading to a nonzero bulk viscosity
  that can impact dynamics and GW signals. We present the first
  simulations of binary neutron-star mergers employing the
  self-consistent and second-order formulation of the equations of
  relativistic hydrodynamics for dissipative fluids proposed by M\"uller,
  Israel and Stewart. With the spirit of obtaining a first assessment of
  the impact of bulk viscosity on the structure and radiative efficiency
  of the merger remnant we adopt a simplified but realistic approach for
  the viscosity, which we assume to be determined by direct and modified
  Urca reactions and hence to vary within the stars. At the same time, to
  compensate for the lack of a precise knowledge about the strength of
  bulk viscosity, we explore the possible behaviours by considering three
  different scenarios of low, medium, and high bulk viscosity. In this
  way, we find that large values of the bulk viscosities damp the
  collision-and-bounce oscillations that characterize the dynamics of the
  stellar cores right after the merger. At the same time, large
  viscosities tend to preserve the $m=2$ deformations in the remnant,
  thus leading to a comparatively more efficient GW emission and to
  changes in the post-merger spectrum that can be up to $100\,{\rm Hz}$
  in the case of the most extreme configurations. Overall, our
  self-consistent results indicate that bulk viscosity increases the
  energy radiated in GWs soon after the merger by $\lesssim 2\%$ in the
  (realistic) scenario of small viscosity, and by $\lesssim 30\%$ in the
  (unrealistic) scenario of large viscosity.
\end{abstract}


\maketitle

\noindent\emph{Introduction.~}The merger of binary neutron stars (BNSs)
is a unique phenomenon in which all of the four fundamental interactions
play, at different stages in the evolution, an important role imprinted
in the astronomical observables measured in different channels. This very
property makes their accurate modelling a challenging multi-physics
problem that however has the potential of providing constraints on
gravity or the strong interaction of dense matter. For example, the first
multi-messenger observation of the BNS merger
GW170817~\cite{Abbott2017_etal, Drout2017, Cowperthwaite2017c} resulted
in numerous bounds on the properties of isolated nonrotating neutron
stars \eg their maximum mass $M_{_{\mathrm{TOV}}}$ or their distribution
in radii~\cite{Annala2017, Bauswein2017b, Margalit2017, Radice2017b,
  Rezzolla2017, Ruiz2017, Most2018, Shibata2019, Annala2019,
  Dietrich2020}, which can be directly used to constrain the equation of
state (EOS) of cold nuclear matter. While many of these results stem from
the gravitational-wave (GW) signal during the inspiral, the post-merger
signal promises to provide even more information about the EOS at extreme
densities \cite{Stergioulas2011b, Bauswein2012a, Takami2014,
  Bernuzzi2015a, Rezzolla2016, Breschi2022a}, especially when considering
the possible appearance of a phase transition to quark
matter~\cite{Most2018b, Most2019c, Weih:2019xvw, Tootle2022,Bauswein2019,
  Blacker2020, Liebling2021, Prakash:2021wpz, Fujimoto:2022c,
  Ujevic2023}. Of particular interest here are those studies that have
recently explored the possibility that the post-merger signal can also
contain signatures of a bulk viscosity generated by violations of weak
chemical equilibrium and resulting from the out-of-equilibrium dynamics
that accompanies the post-merger remnant in the first few milliseconds
since its formation~\cite{Alford2010, Alford2018, Alford:2018lhf,
  Alford2021a, Hammond:2021vtv, Celora2022, Most2022_a, Most2022,
  Espino2024b} (see~\cite{Ghosh2023,Ripley2023,Ripley2024} for some
recent work on the inspiral).

Indeed, several studies of the microphysics of the Urca processes
relative to \textit{npe} or \textit{npe}$\mu$ matter suggest that, right
after the merger, bulk viscosity could be strong enough to damp the
$\mathrm{kHz}$-density oscillations over a timescale $\lesssim
100\,\mathrm{ms}$ after the merger~\cite{Baiotti2016, Alford2020,
  Alford2022a}. In addition, recent investigations exploring the impact
of \textit{npe} Urca processes in BNS simulations have shown that the
bulk-viscous regime is attained in large parts of the hypermassive
neutron star (HMNS)~\cite{Most2022} and that small differences in the GW
signal appear \cite{Hammond2023a}. At the same time, simulations
employing a moment-based scheme to model neutrino transport have not
found evidence for significant out-of-thermodynamic equilibrium effects
as those needed to produce an effective bulk viscosity~\cite{Radice2022}.

Given these contrasting results and suggestions, it is clear that an
accurate assessment of the role played by bulk viscosity cannot be
achieved until fully general-relativistic simulations of BNS mergers are
performed with the inclusion of the dissipative contributions of bulk
viscosity. These require a self-consistent hydrodynamical treatment of
non-perfect fluids, such as that offered by the hyperbolic and
second-order theory by M\"uller, Israel and Stewart
(MIS)~\cite{Mueller1967, Israel1979, Rezzolla_book:2013,
  Gavassino2021,Most2021d,Camelio2022_a} (see also~\cite{Camelio2022_b}
for preliminary studies in spherical symmetry and~\cite{Duez2004b,
  Radice2017, Shibata2017a, Shibata2017b, Duez2020} for other interesting
approaches to viscous dissipation in general-relativistic
simulations).

In this \textit{Letter} we present the results of such a self-consistent
approach, thus providing the first mathematically and physically firm
estimates of the role of bulk viscosity on the post-merger GW signal.
More specifically, by adopting a systematic but realistic prescription of
bulk viscosity, we probe the extremes of the possible range of strengths
of bulk viscosity and hence quantify under what conditions bulk-viscous
effects can affect the post-merger GW signal from BNSs, obtaining strict
upper limits.

\noindent\emph{Methods.~}As anticipated, we employ the second-order MIS
framework to model dissipative deviations from the perfect-fluid dynamics
(see also~\cite{Disconzi2017, Bemfica2019b, Kovtun2019, Hoult2020,
  Taghinavaz2020, Pandya2022b, Bantilan2022} for recent progress on
first-order theories). In this framework, and assuming that bulk
viscosity is the only relevant dissipative effect on the timescale of
tens of milliseconds or less (see~\cite{Alford2018} for a discussion
about why shear viscosity operates on much longer timescales), the
energy-momentum tensor simplifies considerably and the evolution equation
for the bulk-viscous pressure is given by (see also~\cite{Chabanov2021}
for details)
\begin{align}
  \label{eq:bulk}
  \tau u^{\mu}\nabla_{\mu}\Pi=-\zeta\Theta -\Pi\,,
\end{align}
where $\boldsymbol{u \cdot \nabla} = u^{\mu} \nabla_{\mu}$ refers to the
covariant derivative along the fluid four-velocity $\boldsymbol{u}$ and
the quantities $\Pi$, $\tau$, $\zeta$, and $\Theta :=\nabla_{\mu}
u^{\mu}$ denote the bulk-viscous pressure, the relaxation time, the bulk
viscosity and the fluid expansion, respectively. This approach is
normally referred to as the ``Maxwell-Cattaneo'' model
\cite{Camelio2022_a} since Eq.~\eqref{eq:bulk} can be recovered through
linearisation in out-of-equilibrium contributions, \ie in $\Pi$, of the
more complete Hiscock-Lindblom model presented in
Ref.~\cite{Hiscock1983}. The Hiscock-Lindblom model is based on an
extension of the perfect-fluid entropy current where all possible terms
that depend linearly and quadratically on the dissipative currents are
included. Positivity of entropy production then yields relaxation-type
equations similar to \eqref{eq:bulk} which is characteristic of the MIS
theory. However, it was found that both models are in good agreement
with the full chemically reacting multi-fluid model in simulations of
isolated neutron stars in spherical symmetry~\cite{Camelio2022_b}.

Being the first study of this type, it lacked a previous, even
qualitative, understanding of the impact that bulk viscosity may have on
the dynamics and GW emission after the merger. To build such
understanding, we have adopted an approximate but realistic description
of bulk viscosity considering it to be determined by direct and modified
Urca reactions using the linearized equations presented in
\cite{Camelio2022_a, Camelio2022_b}. Hence, both $\zeta$ and $\tau$
  are sensitive functions of the baryon-number density $n$ and of the
  temperature $T$, varying within the stars depending on the local
  thermodynamic conditions, \eg the bulk viscosity associated with the
  linearized direct Urca reaction rate scales as $\sim T^{4}$. However,
its strength is systematically varied by adjusting the composition of
cold neutron-star matter above the nuclear saturation density (see End
Matter for details on the EOS and the transport coefficients
\cite{Chabanov2023endmatter}). The advantage of this approach is that it
allowed us to explore the most extreme scenarios of very large and very
small bulk viscosities. In practice, we consider four different models of
the bulk-viscosity which can be distinguished by computing the resonant
maximum of the so-called AC bulk viscosity \cite{Yang2023}
$\zeta_{_{\mathrm{AC}}}(\omega) := \zeta(1 + \omega^2\tau^2)^{-1}$, where
$\omega$ is the angular frequency of a periodic density oscillation
\cite{Alford2018a, Sawyer1989}. The quantities $\hat{\zeta}$ and
$\hat{T}$ define the resonant maximum of $\zeta_{_{\mathrm{AC}}}$ at
$2n_{\mathrm{sat}}$ and a frequency $f=\omega/2\pi=1~\mathrm{kHz}$, where
$n_{\mathrm{sat}}\approx0.15~\mathrm{fm}^{-3}$ is the nuclear saturation
density, and the corresponding temperature at which the maximum is
achieved, respectively. Our models span the range of $\hat{\zeta} \in
\zeta_0 [10^{-4},0.4, 1.0, 2.4]$ with $\zeta_0 := 10^{30}
\mathrm{g}\,\mathrm{cm}^{-1}\,\mathrm{s}^{-1}$ and correspondigly
$\hat{T} \in [3.6,2.2,1.3,1.2]~\mathrm{MeV}$. Hereafter, we will refer to
these four cases as: ``zero'' (because of its negligible value), low,
medium, and high viscosity, respectively.

\begin{figure*}
\includegraphics[width=0.33\textwidth]{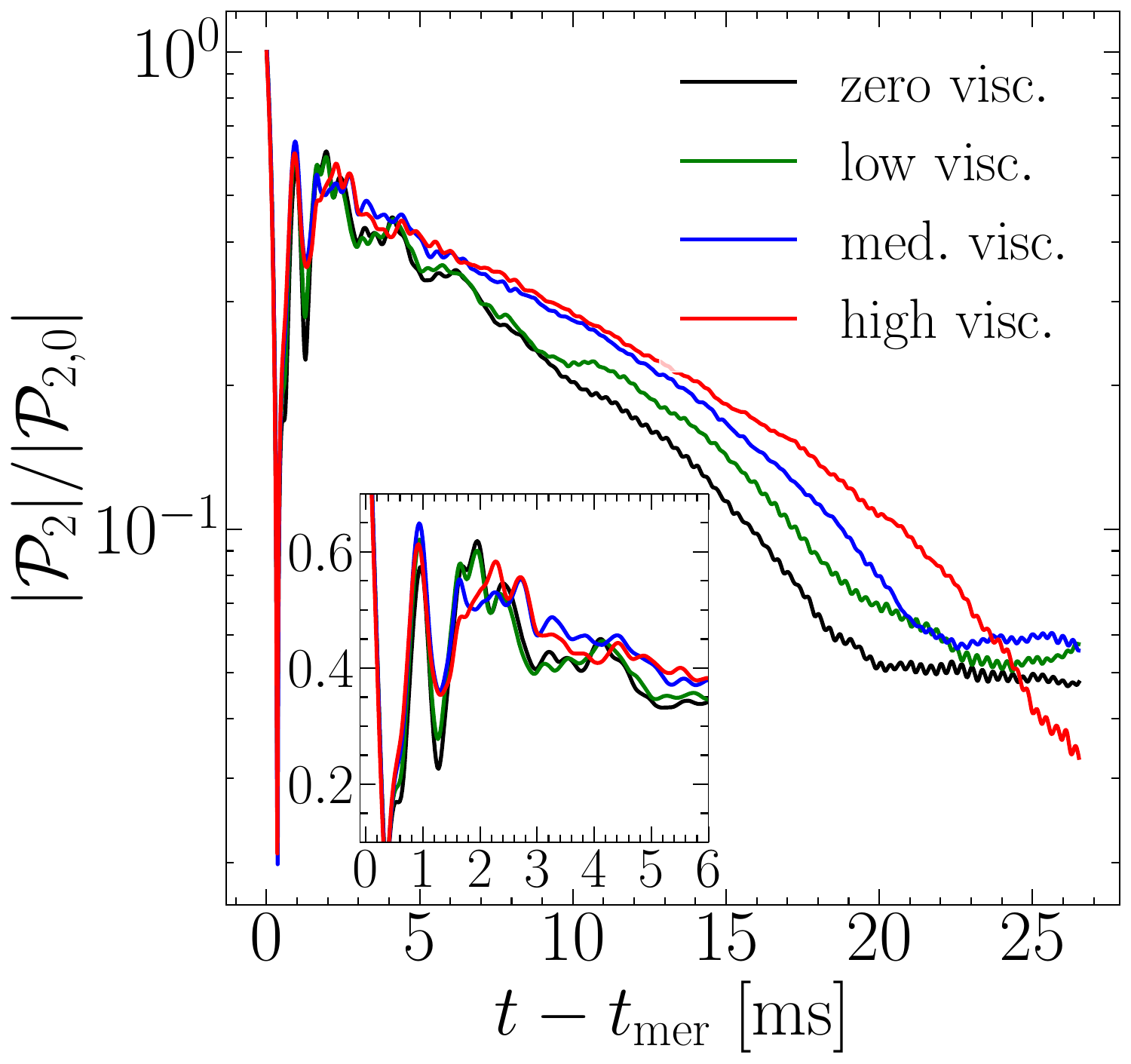}
\hskip 0.1cm
\includegraphics[width=0.33\textwidth]{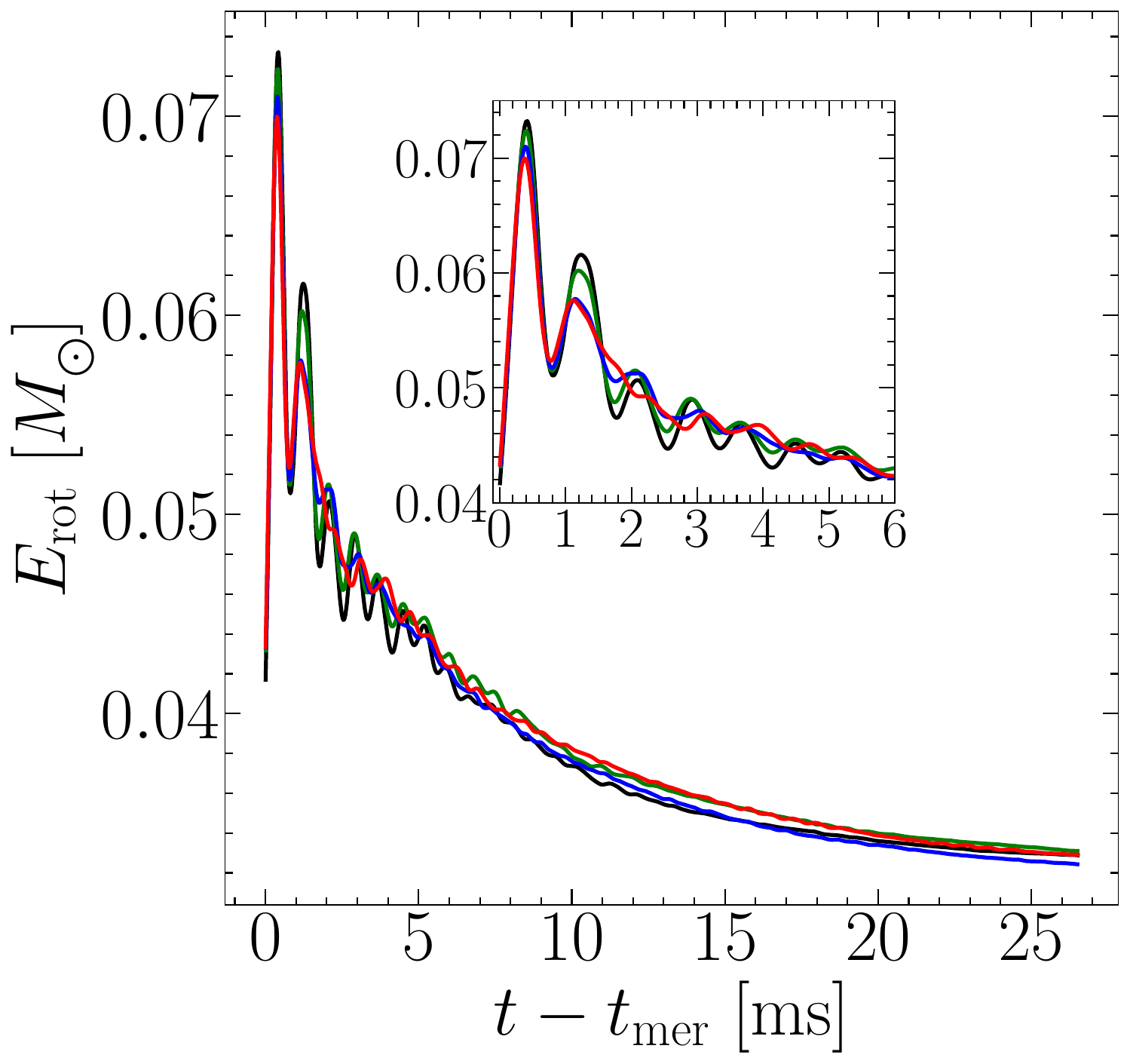}
\hskip 0.1cm
\includegraphics[width=0.32\textwidth]{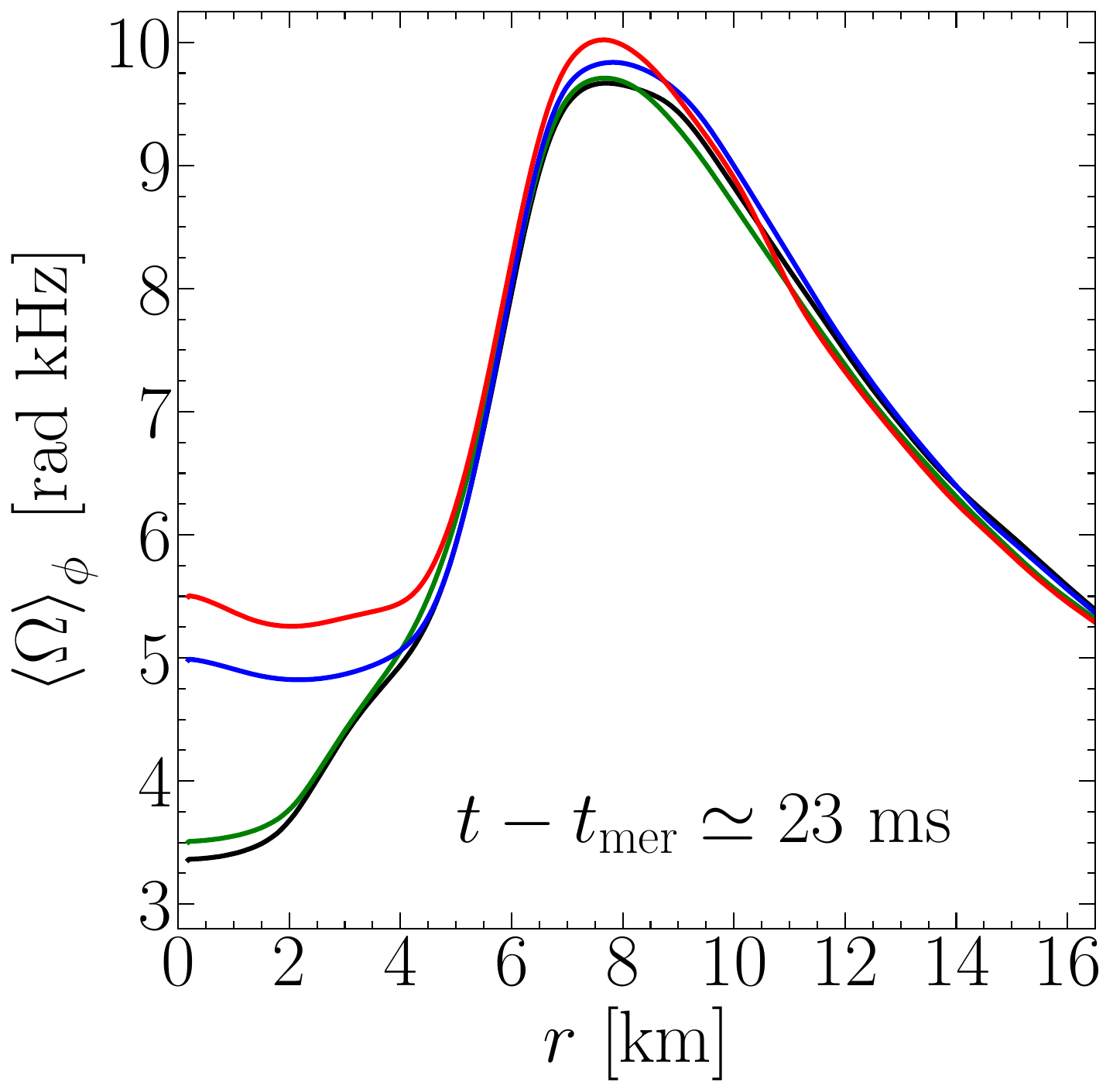}
\caption{\textit{Left:} Evolution of the $m=2$ rest-mass density mode
  $\mathcal{P}_{2}$ normalized to its value at the time of the merger
  $\mathcal{P}_{2,0}$ for the four configurations
  considered. \textit{Middle:} The same as on the left but for the
  rotational kinetic energy. \textit{Right:} $\phi$-averages in the
  equatorial plane of the angular velocity as a function of the
  coordinate radius $r$ at a representative late time
  ($t-t_{\mathrm{mer}}\approx 23\,\mathrm{ms}$).}
  \label{fig:mtwo}
\end{figure*}

The simulations reported below are obtained after solving the Einstein
equations together with an extension of the high-order high-resolution
shock-capturing code \texttt{FIL}~\cite{Most2019b, Most2019c} and an
equilibrated hybrid EOS, \ie where the pressure is expressed as a
combination of a cold part modelled via a $\beta$-equilibrium slice of
the \texttt{TNTYST} EOS~\cite{Togashi2017} and of a thermal part
described by an ideal-fluid EOS~\cite{Rezzolla_book:2013}. In essence, we
set the pressure as $p^{\mathrm{eq}}=p_{\mathrm{cold}}+\rho
\epsilon_{\mathrm{th}} (\Gamma_{\rm th} -1)$, where $\rho$ is the
rest-mass density and $\epsilon_{\mathrm{th}}$ the thermal part of the
specific internal energy. We also adopt an optimal value of the adiabatic
index $\Gamma_{\rm th}=1.7$~\cite{Figura2020}. Note that only
$p^{\mathrm{eq}}$ is employed in the simulations as deviations from
equilibrium are modelled through $\Pi$. As mentioned above, we vary the
composition of cold neutron-star matter to obtain different models for
the bulk viscosity while keeping the same $p^{\mathrm{eq}}$ for all
simulations.

The \texttt{FIL} code employs fourth-order accurate finite-difference
stencils in Cartesian coordinates for the evolution of the constraint
damping formulation of the Z4 formulation of the Einstein
equations~\cite{Bernuzzi:2009ex, Alic:2011a}, while the equations of
relativistic magnetohydrodynamics are solved with a fourth-order
high-resolution shock-capturing scheme~\cite{DelZanna2002, DelZanna2007}
(for simplicity, we consider zero magnetic fields). In particular, we
solve Eq.~\eqref{eq:bulk} following the strategy presented in
Ref.~\cite{Chabanov2021, Chabanov2023b}, making sure that for small
densities, \ie $\rho < \rho_{\rm th}\approx 4.5\times 10^{14}
\mathrm{g}\,\mathrm{cm}^{-3}$, $(\zeta/\tau)$ and $\tau$ are interpolated
between their microphysical values and their atmosphere values via a
power law. We also explicitly enforce causality by adjusting
$(\zeta/\tau)$ dynamically when causality is violated using the
expression suggested by Ref.~\cite{Bemfica2019}, where the full nonlinear
characteristic velocity is set to $0.99$. Additionally, we employ the
same limiting procedure introduced in \cite{Chabanov2023b}, see
Section~\rom{2}.B.1, to avoid unphysically large values of the
bulk-viscous pressure by imposing $-0.9\, p^{\mathrm{eq}} \leq \Pi \leq
e-p^{\mathrm{eq}}$, where $e$ is the energy density. Finally, we impose a
floor on $\tau$ in order to ensure numerical stability using $\tau\geq
1.1\Delta t_{l}$, where $\Delta t_{l}$ denotes the timestep on a given
refinement level $l$.

The initial data is computed as in \cite{Chabanov2022} using the
\texttt{FUKA} code~\cite{Papenfort2021b}, where the equal-mass binaries
are chosen to be irrotational with a total ADM mass of $\sim
2.55~M_{\odot}$ at a separation of $\sim 30~M_{\odot}\approx
44\,\mathrm{km}$. The computational grid has outer boundaries at
$1000\,M_{\odot}\simeq 1476\,{\rm km}$ in the three spatial directions
and we employ a $z$-symmetry across the equatorial plane. All simulations
have been performed with a reference resolution of $\Delta x\sim
0.17~M_{\odot}\approx 260\,\mathrm{m}$ on the sixth refinement level;
however, additional simulations have been performed at a lower resolution
of $\Delta x\sim 0.25~M_{\odot}\approx 370\,\mathrm{m}$ for all
scenarios. The results obtained at different resolutions are
mathematically consistent and physically robust.

\noindent\emph{Results.~}We start by describing the overall evolution of
the binaries focussing on the structural and the rotational properties of
the merger remnant, which is represented by an HMNS. A convenient manner
to measure the differences in the HMNS structure is offered by a Fourier
decomposition of the rest-mass density~\cite{Baiotti06b, Franci2013,
  Radice2016a}
\begin{align}
\mathcal{P}_{m} :=\int \rho W e^{-im\phi}\, \sqrt{\gamma}\, dxdydz\,,
\label{eq:mode}
\end{align}
where $W$ is the Lorentz factor, $\gamma$ the determinant of the spatial
metric and $\phi$ the azimuthal angle.

The left panel of Fig.~\ref{fig:mtwo} reports the evolution of the $m=2$
density mode, or bar-mode, $\mathcal{P}_{2}$ when normalized to its value
at the time of the merger $\mathcal{P}_{2,0}$, and for the four models of
the bulk viscosity considered in our simulations. Note that the initial
oscillations for $t-t_{\rm mer} \lesssim 3\,{\rm ms}$ are the result of
the rapid and quasi-periodic collisions of the two stellar cores
(see~\cite{Takami2015} for a mechanical toy model). Furthermore, the
evolution of $\mathcal{P}_{2} / \mathcal{P}_{2,0}$ is impacted in a
systematic manner by the strength of the bulk viscosity, which
``preserves'' the initial $m=2$ deformation of the binary and hence leads
to a less axisymmetric HMNS. This result is the effect of large and
anisotropic temperature gradients inside the HMNS remnant, see also
\cite{Hanauske2016} regarding the occurence of so-called temperature
``hot spots'', because of the strong temperature dependence of the
considered reaction rates. These temperature anisotropies are dominated
by a similar $m=2$ deformation sourced by the relatively cold cores (see
End Matter). Since the bar-deformation represents a way in which the HMNS
minimises its rotational kinetic energy~\cite{Baiotti06b}, it is
interesting to evaluate the rotational properties of the HMNS.

The middle panel of Fig.~\ref{fig:mtwo} reports the evolution of the
rotational kinetic energy [see, \eg Eq.~(12.48)
  of~\cite{Rezzolla_book:2013} for a definition] and highlights three
important effects. First, the initial oscillations in $E_{\mathrm{rot}}$
are in phase opposition to those in $\mathcal{P}_{2}$ since the bar-mode
deformation is reduced when the two cores have the smallest separation
and the HMNS has the largest rotational kinetic energy. Second, these
oscillations have a damping that is stronger as the bulk viscosity is
increased and this can again be easily understood in terms of the toy
model, where the bulk viscosity plays the role of a friction in the
mechanical oscillator (see inset). Finally, the overall evolution of the
rotational kinetic energy is independent of the strength of the bulk
viscosity. To understand this result it is important to bear in mind that
two processes are at play at the same time. First, by enhancing
non-axisymmetric deformations, bulk viscosity is effectively increasing
GW emission and, hence, more efficiently leading to removal of rotational
kinetic energy and angular momentum from the HMNS. Second, GW emission
effectively acts as a torque on the fluid, leading to fluid elements
moving inward and attaining a higher angular velocities. As a result, the
HMNS becomes more compact and spins-up, converting the lost rotational
kinetic energy into gravitational binding energy. However, these changes
are large only when $\hat{\zeta} \simeq \zeta_0$, being much smaller and
almost negligible for $\hat{\zeta} \ll \zeta_0$.

\begin{figure*}
\includegraphics[width=0.295\textwidth]{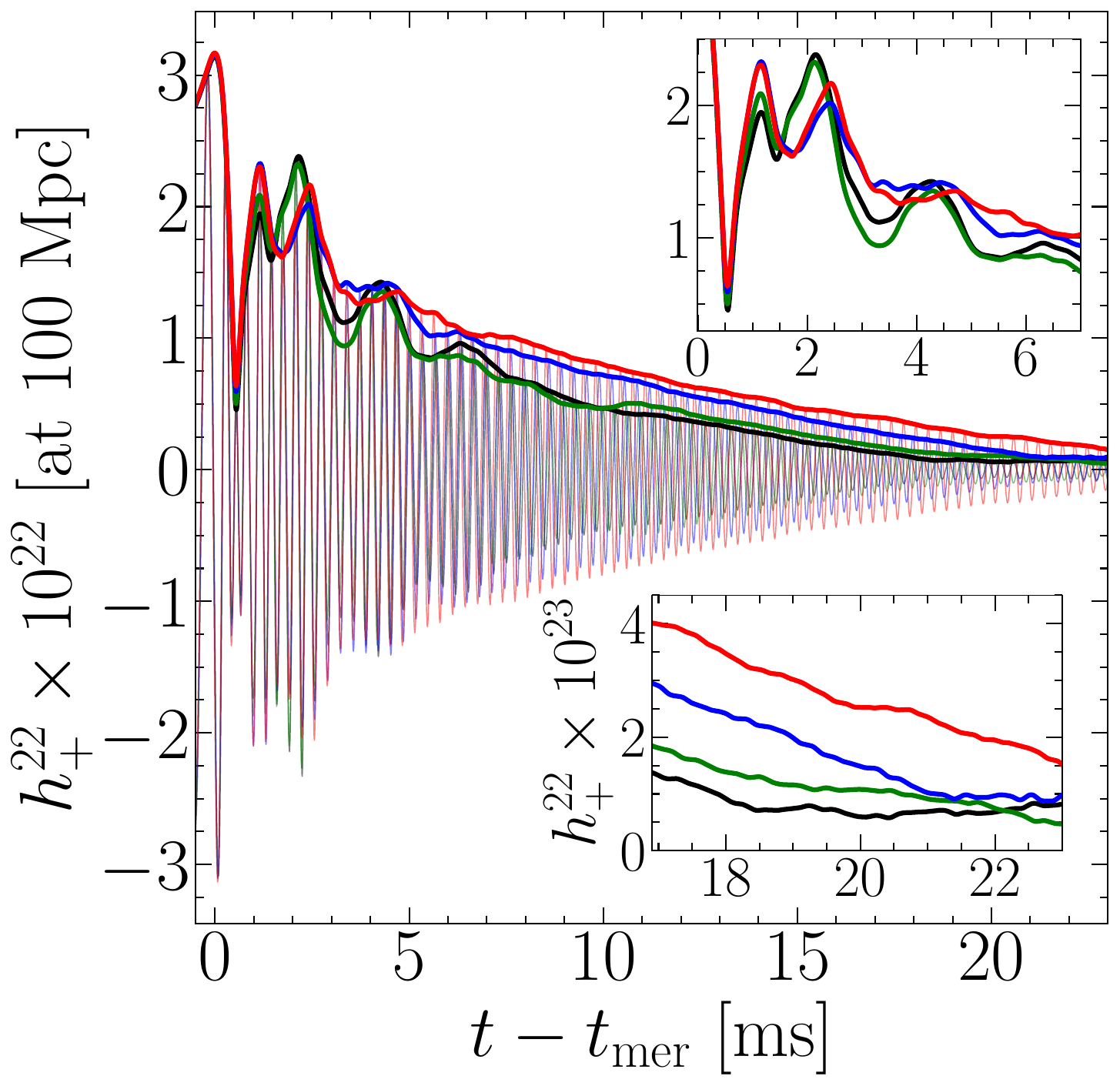}
\hskip 0.1cm
\includegraphics[width=0.315\textwidth]{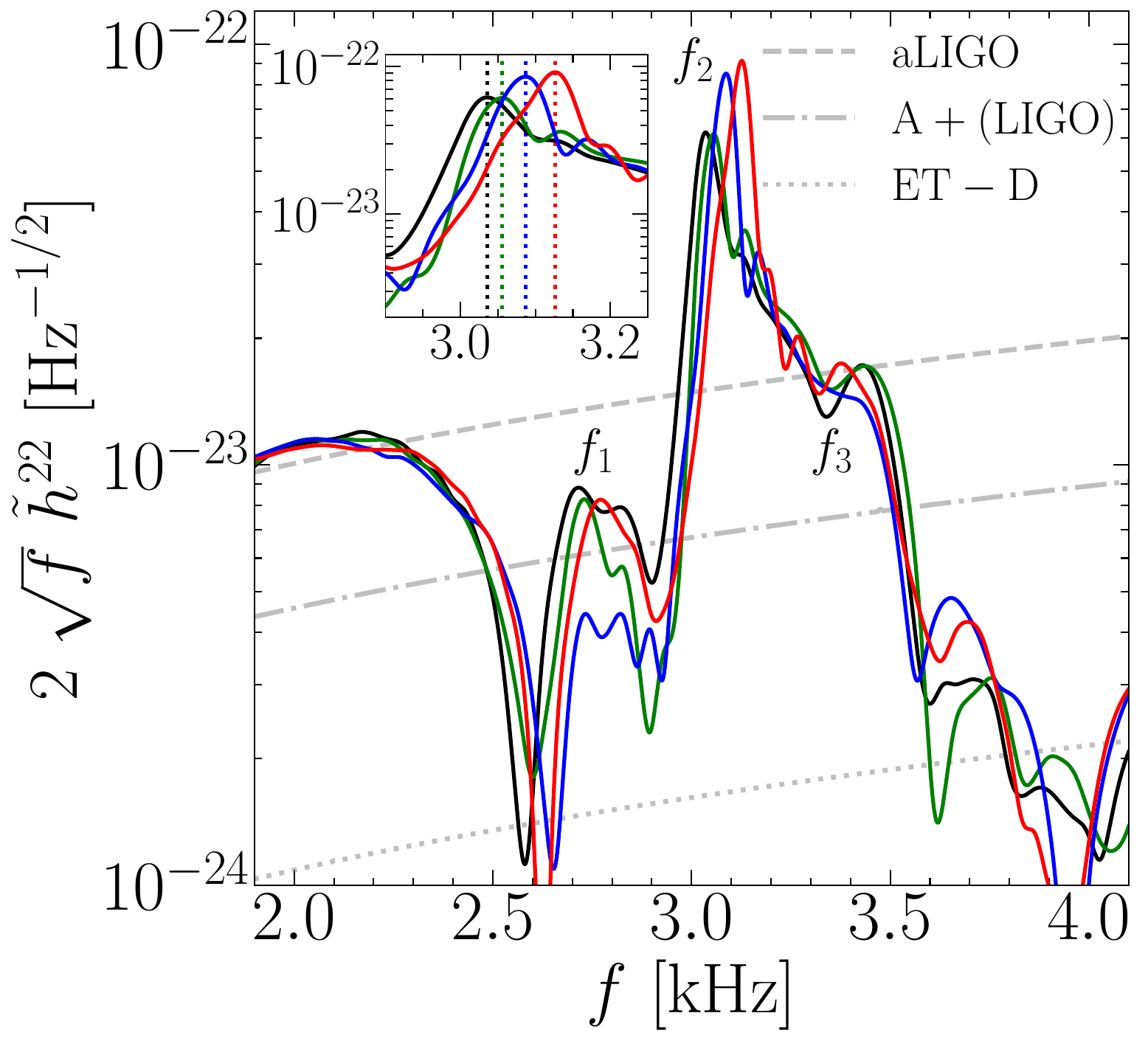}
\hskip 0.1cm
\includegraphics[width=0.362\textwidth]{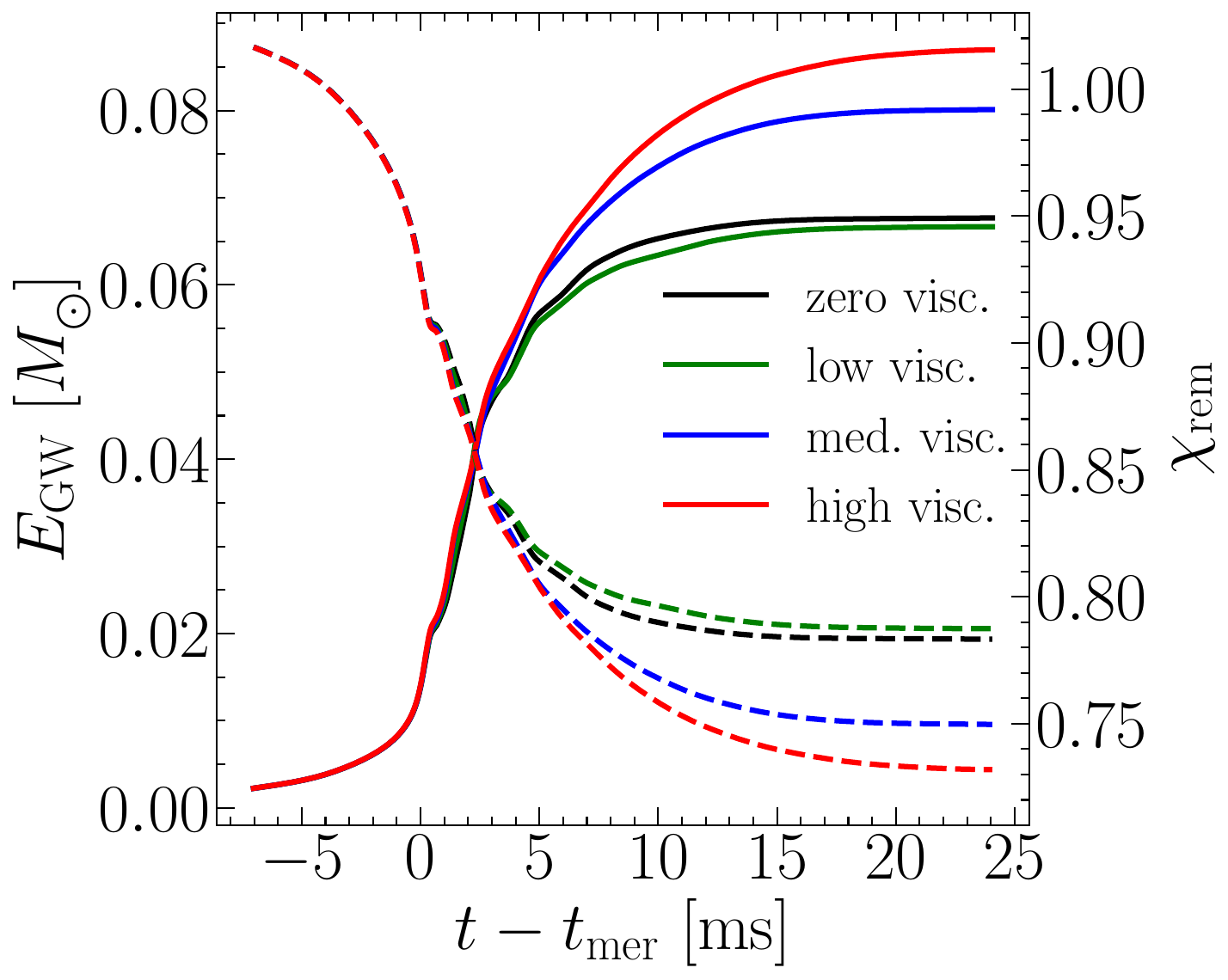}
\caption{\textit{Left:} GW strain in the $\ell=m=2$ mode of the
  $+~$--polarization extracted at $\sim 740\,\mathrm{km}$ and normalized
  to a distance of $100\,{\rm Mpc}$ for the four configurations
  considered. Thick solid lines report the corresponding amplitudes.
  \textit{Middle:} Post-merger PSD of the GW signals on the left. Dashed,
  dash-dotted and dotted lines report sensitivity curves for advanced
  LIGO (aLIGO), the A+ configuration, and the Einstein Telescope,
  respectively~\cite{Punturo:2010, Cahillane2022}), detectors.
  \textit{Right:} Evolution of the radiated GW energy (solid lines) and
  of the dimensionless spin of the HMNS remnant (dashed lines). Both
  quantities are measured at $\sim 740\,\mathrm{km}$. }
\label{fig:waves}
\end{figure*}

Further evidence for this dynamics is provided in the right panel of
Fig.~\ref{fig:mtwo}, which shows the azimuthal ($\phi$) averages in the
equatorial plane of the angular velocity as a function of the coordinate
radius $r$ at a representative late time, \ie $t-t_{\mathrm{mer}} \approx
23\,\mathrm{ms}$ (time averages over $2\,$ ms yield very similar
profiles). The right panel of Fig.~\ref{fig:mtwo} clearly shows that the
angular velocity of the HMNS core, \ie for $r\lesssim 8\,\mathrm{km}$ is
larger in the case of large bulk viscosities and after reaching a local
maximum at $\simeq 7-8\,{\rm km}$, it falls off following a Keplerian
profile. 

The influence of bulk viscosity on the GW signal is summarised in
Fig.~\ref{fig:waves}, which reports the post-merger GW signal (left
panel), the corresponding power spectral density (PSD, middle panel) and
with solid lines the radiated energy $E_{\mathrm{GW}}$ (right panel).
Following \cite{Zhu2021}, we calculate the PSD by using only the GW
signal in the time window $t - t_{\rm mer} \in [-7,23]~\mathrm{ms}$ (\ie
from about $\sim 3.5$ orbits before merger) and a Tukey window function
with $\alpha=0.25$. Overall, the amplitude of the $\ell=2,m=2$ GW mode is
larger for higher bulk viscosities as anticipated from the evolution of
the $m=2$ deformation of the binary. Additionally, we observe a dephasing
between the various waveforms which starts at $t-t_{\mathrm{mer}}\approx
5\,\mathrm{ms}$ and can be attributed to the different evolutions of the
maximum density.

For $t-t_{\mathrm{mer}}\lesssim 5\,\mathrm{ms}$, the binary undergoes
very violent collision-and-bounce cycles which rapidly increase the
maximum density. Even though larger bulk viscosities lead to a stronger
damping of these collision-and-bounce cycles while preserving the $m=2$
deformations, the maximum rest-mass densities at
$t-t_{\mathrm{mer}}\approx 5\,\mathrm{ms}$ are comparable across all
values of the viscosity. However, for $t-t_{\mathrm{mer}}\gtrsim
5\,\mathrm{ms}$, the growth of the maximum rest-mass density becomes
significantly slower as the kinetic energy responsible for the
collision-and-bounce cycles has been largely dissipated. At this point,
the evolution of the maximum rest-mass density starts to differ for
different viscosities, with larger viscosities leading to larger maximum
densities, reflecting the increased loss of angular momentum due to GW
emission.

While the signal-to-noise ratio of these post-merger GW signals is only
moderately increased by viscosity (the largest differences are of $\sim
10\%$), the most important impact that bulk viscosity has on the GW
signal is the systematic shift to higher frequencies of the largest peak
of the PSD, $f_{2}$, with differences that can be of $\simeq 4\%$ or
$\simeq 100\,\mathrm{Hz}$. High-frequency shifts of this type have been
measured when considering the changes induced by magnetic fields (see,
\eg~\cite{Giacomazzo2011b}) and are comparable with the precision with
which this frequency is employed via quasi-universal relations (see, \eg
\cite{Bauswein2012, Takami2014, Bernuzzi2014, Takami2015, Rezzolla2017})
(comparable shifts to higher frequencies have been observed in
Refs.~\cite{Hammond2023a, Most2022_a}). This shift can be easily
understood by keeping in mind that higher viscosities lead to larger
bar-mode deformations and larger spinning frequencies, hence to an $f_2$
frequency that is both larger in amplitude and at higher
frequency. Clearly, this viscosity-driven frequency shift needs to be
accounted for when estimating the error budget in the quasi-universal
relations for $f_2$. At the same time, it reveals an additional
degeneracy with the changes induced by magnetic fields and thermal
effects, thus impacting on the precision with which the EOS can be
determined from the post-merger spectrum.

As a concluding remark, we discuss the impact that bulk viscosity has on
the amount of radiated GW energy and angular momentum. The right panel of
Fig.~\ref{fig:waves} shows with solid lines the radiated energy
$E_{\mathrm{GW}}$ and it is straightforward to appreciate that larger
bulk viscosities lead to a larger energy loss. Also, while the radiated
energies differ by $\lesssim 2\%$ for the low-viscosity binary with
$\hat{\zeta} = 0.4\,\zeta_0$, the difference can be as large as $\simeq
30\%$ when $\hat{\zeta} = 2\,\zeta_0$\footnote{Also in this case, the
behaviour of $E_{\mathrm{GW}}$ is not monotonic for small bulk
viscosities and at the reference resolution.}. A very similar behaviour
is also shown by the dimensionless spin of the HMNS remnant, which we
define as~\cite{Papenfort:2022ywx}: $\chi_{\mathrm{rem}} :=
({J_{\mathrm{ADM}} - J_{\mathrm{GW}}})/({M_{\mathrm{ADM}} -
E_{\mathrm{GW}})^{2}}$, where $M_{\mathrm{ADM}}$ and $J_{\mathrm{ADM}}$
are the initial ADM mass and angular momentum, respectively, and
$J_{\mathrm{GW}}$ is the $z$-component of radiated angular momentum. The
behaviour of $\chi_{\mathrm{rem}}$ is reported with dashed lines in the
right panel of Fig.~\ref{fig:waves} and shows that bulk viscosity will
impact the spin of the black hole if the HMNS spins down considerably and
collapses, which could be up to $\sim 8\,\%$ smaller.

\noindent\emph{Conclusions.~}To obtain a mathematically consistent and
quantitatively robust assessment of the role played by bulk viscosity in
BNS mergers, we have carried out the first simulations of merging
binaries where dissipative effects are accounted for self-consistently
within the causal and second-order formulation of dissipative
hydrodynamics by MIS. Our microphysical description consists of a bulk
viscosity that is determined by direct and modified Urca reactions which
are responsible for achieving weak chemical equilibrium in neutron-star
matter. In addition, we vary the composition of cold neutron-star matter
above the nuclear saturation density in order to study systematically the
impact of small (realistic) and large (unrealistic) bulk viscosities. We
have explored four values of the resonant maximum bulk-viscosity
coefficient at $2\,n_{\mathrm{sat}}$ and $1~\mathrm{kHz}$, \ie
$\hat{\zeta} \in \zeta_0[10^{-4}, 0.4, 1.0, 2.4]$, where $\zeta_0 :=
10^{30}\,\mathrm{g}\,\mathrm{cm}^{-1}\, \mathrm{s}^{-1}$.

Overall, our study reveals that large bulk viscosities are effective at
damping the collision-and-bounce oscillations of the stellar cores while,
at the same time, preserving the initial $m=2$ deformations of the
binary. Second, the increase of the bar-mode deformation increases the
efficiency of energy and angular-momentum losses via GWs. Third, the
stellar structure of the HMNS is modified and is characterised by a more
compact remnant with uniformly rotating core spinning faster than in the
inviscid case but having the same rotational kinetic energy. Finally, and
more importantly, the larger spinning frequency of the viscous remnants
is reflected in a larger value for the $f_2$ frequency in the post-merger
PSD, which needs to be properly taken into account to infer the
properties of the EOS via universal relations (see, \eg
\cite{Bauswein2012a, Bauswein2014, Bernuzzi2015a, Bauswein2015b,
  Bose2017, Lioutas2021, Breschi2022a}. While the behaviour described
above applies to all viscous binaries, the differences between viscous
and inviscid binaries become significant only for $\hat{\zeta} \gtrsim
\zeta_{0}$. As a result, our self-consistent results indicate that
bulk-viscous effects increase the radiated energy by $\lesssim 2\%$ in
the (realistic) scenario of small viscosity, and at most by $\sim 30\%$
in the (unrealistic) scenario of large viscosity.

The work presented here can be improved in a number of ways. First and
foremost, by employing a full temperature-dependent EOS and together with
the proper inclusion of the modifications introduced by neutrino emission
and absorption, we would be able to have a more realistic model for the
temperature profile of the HMNS remnant. Second, a more extensive
exploration of different EOSs is necessary to remove a possible bias in
our conclusions, which may arise from investigating only one EOS. Third, taking into
account the presence of trapped neutrinos in the HMNS remnant
\cite{Most2022,Espino2024b} would allow for a more realistic composition
of hot neutron-star matter in the post-merger. We plan to explore these
aspects in future work.

\medskip
\begin{acknowledgments}
  It is a pleasure to thank M. Alford, M. Hanauske, E. Most, and
  K. Schwenzer for useful comments and discussions. We are also grateful
  to the Referees, who have pushed us to consider a non-constant and more
  description of bulk viscosity. Partial funding comes from the GSI
  Helmholtzzentrum f\"ur Schwerionenforschung, Darmstadt as part of the
  strategic R\&D collaboration with Goethe University Frankfurt, from the
  State of Hesse within the Research Cluster ELEMENTS (Project ID
  500/10.006), by the ERC Advanced Grant ``JETSET: Launching, propagation
  and emission of relativistic jets from binary mergers and across mass
  scales'' (Grant No. 884631) and the Deutsche Forschungsgemeinschaft
  (DFG, German Research Foundation) through the CRC-TR 211
  ``Strong-interaction matter under extreme conditions''-- project number
  315477589 -- TRR 211. LR acknowledges the Walter Greiner Gesellschaft
  zur F\"orderung der physikalischen Grundlagenforschung e.V. through the
  Carl W. Fueck Laureatus Chair. The simulations were performed on HPE
  Apollo HAWK at the High Performance Computing Center Stuttgart (HLRS)
  under the grant BNSMIC. MC acknowledges support from the NSF grants
  PHY-2110338, PHY-2409706, AST-2031744 and OAC-2004044 as well as the
  NASA TCAN Grant No. 80NSSC24K0100.
\end{acknowledgments}

\bibliography{bulk_vis_gws_letter}

\appendix
\noindent\section{End Matter}

In what follows we provide the most important details about the
calculations or considerations that are needed to obtain the results
presented in the main text.

\noindent\emph{Equations of state.~}
\label{Sec:3+1}
First, in the following we set the Boltzmann constant
$k_{\mathrm{B}}=m_{\mathrm{b}},$ where $m_{\mathrm{b}}$ denotes the
baryon mass. Our models for the EOS closely resemble those employed in
\cite{Camelio2022_a, Camelio2022_b}.

The total specific internal energy of our model is a sum of three
components, \ie $ \epsilon = \epsilon(\rho,T,Y_e) =
\epsilon_{\mathrm{cold}} + \epsilon_{\mathrm{th}} + \epsilon_e$, with
\begin{align}
\epsilon_{\mathrm{cold}}&=\epsilon_{\mathrm{cold}}(\rho)\,,\\
\epsilon_{\mathrm{th}}&=\left(\Gamma_{\mathrm{th}}-1\right)^{-1}T=c_{_V}T\,,\\
\epsilon_{e}&=k_e\left(Y_e-Y^{\mathrm{eq}}_e(\rho)\right)^2\,,
\end{align}
where $\epsilon_{\mathrm{cold}}$ is the specific internal energy of a
cold $\beta$-equilibrium slice of the original full temperature and
composition dependent EOS, $k_e$ is a constant parameter, the constant
$c_{_V}=(\Gamma_{\mathrm{th}}-1)^{-1}$ the specific heat at constant
volume of the fluid \cite{Rezzolla_book:2013} and $Y_e^{\mathrm{eq}}$ is
the electron fraction in $\beta$-equilibrium. 

Additionally, we construct the entropy of our model by employing only a
thermal contribution from an ideal gas of baryons by assuming that this
component dominates over the entropy contribution from the cold matter
part, \ie $s=s(\rho,T)=k_{\mathrm{B}}\ln\left[
  m_{\mathrm{b}}\rho^{-1}\Phi^{-1} \left(\epsilon_{\mathrm{th}} /
  c_{_V}\right)^{c_{_V}}\right]$, where $s$ is the entropy per baryon and
$\Phi$ an integration constant \cite{Landau-Lifshitz5-1}.

As a result, the free energy per baryon is given by
$f=f(\rho,T,Y_e)=m_{\mathrm{b}}\epsilon(\rho,T,Y_e)-Ts(\rho,T)$ and we
can fully determine all thermodynamic properties of the system. For
example, the pressure is
$p=\rho^2\left(\mathrm{d}\epsilon_{\mathrm{cold}}/\mathrm{d}\rho\right)
+(\Gamma_{\mathrm{th}}-1)\rho\epsilon_{\mathrm{th}}
-2k_e\rho^2\left(Y_e-Y^{\mathrm{eq}}_e\right)
\left(\mathrm{d}Y^{\mathrm{eq}}/\mathrm{d}\rho\right)\simeq
p^{\mathrm{eq}}-2k_e\rho^2\left(Y_e-Y^{\mathrm{eq}}_e\right)
\left(\mathrm{d}Y^{\mathrm{eq}}/\mathrm{d}\rho\right)$.  Note that in
equilibrium, \ie $Y_e=Y_e^{\mathrm{eq}}$, we recover the equilibrium
hybrid pressure defined in the main text, \ie
$p^{\mathrm{eq}}=p_{\mathrm{cold}}+\left(\Gamma_{\mathrm{th}} -
1\right)\rho\epsilon_{\mathrm{th}}$, which is employed in our
simulations. Also, note that the deviations between $p_{\mathrm{cold}}$
and $\rho^2\mathrm{d}\epsilon_{\mathrm{cold}}/\mathrm{d}\rho$ are
negligible for almost all densities. Only at very low densities
$p_{\mathrm{cold}}$ has significant thermal contributions because it was
obtained from slicing at the lowest available temperature, \ie at
$T=0.1~\mathrm{MeV}$, of the original Table.

Similarly, we find the electron chemical potential to be $\mu_e =
2m_{\mathrm{b}} k_e \left( Y_e-Y_e^{\mathrm{eq}}\right)$. Note that by
construction $\mu_n = \mu_p$ such that the condition for
$\beta$-equilibrium becomes $0=\mu_e$ which is satisfied for
$Y_e=Y_e^{\mathrm{eq}}$. In addition, we compute the affinity
$\mathbb{A}$, which measures deviations from chemical (in our case
$\beta$-) equilibrium \cite{Gavassino2021, Camelio2022_a}, and obtain
$\mathbb{A} = -m_{\mathrm{b}}\left(\partial \epsilon/\partial Y_e
\right)_{\rho,s} = -\mu_e$.

\begin{figure}
\includegraphics[width=0.49\textwidth]{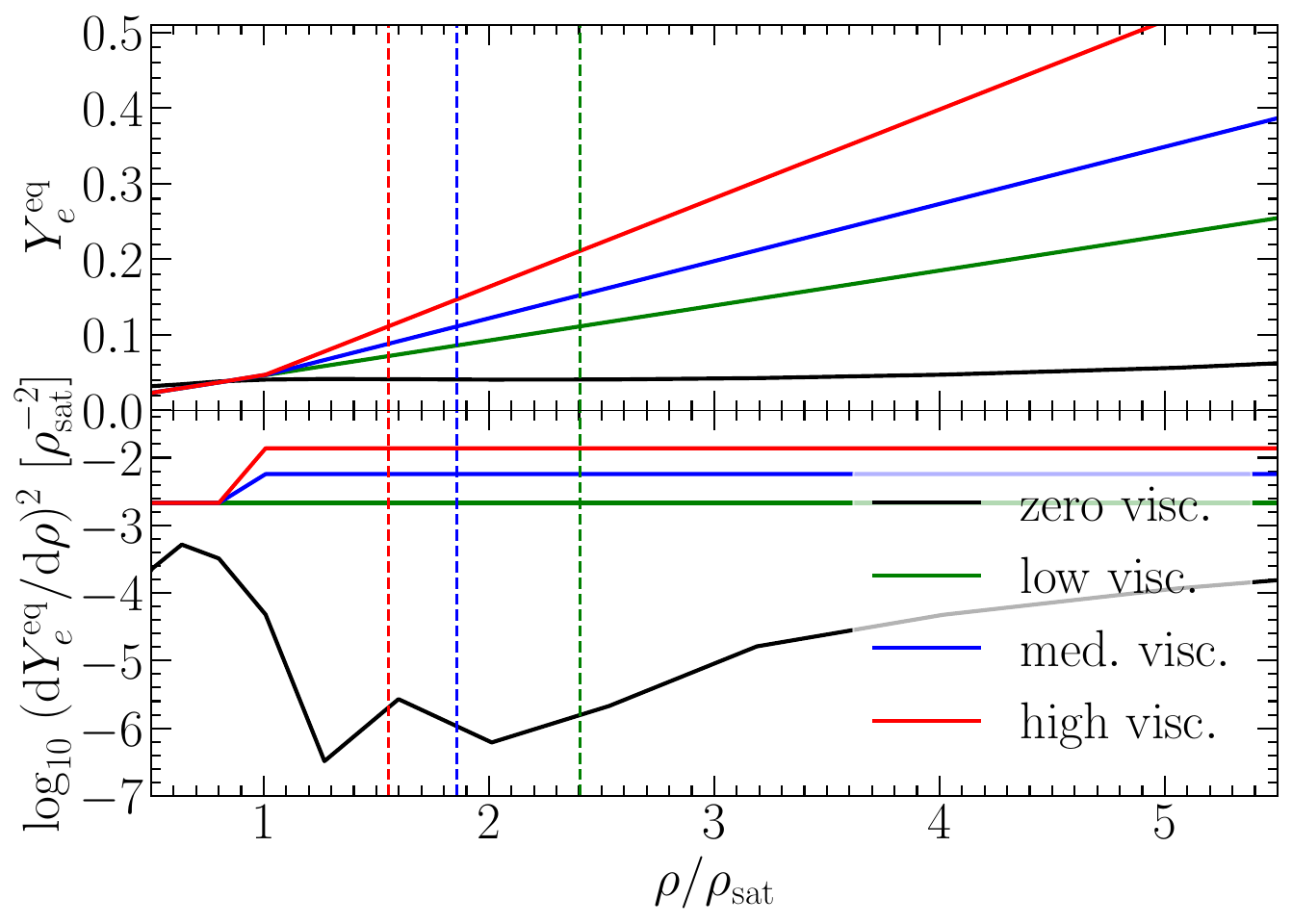}
\caption{\textit{Top:} Electron fraction of $\beta$-equilibrated neutron
  star matter for the zero (black), low (green), medium (blue) and
  high (red) viscosity case. \textit{Bottom:} Square of the derivative
  of the electron fraction in $\beta$-equilibrium. Dashed vertical lines
  show the direct Urca density threshold for each of the cases
  considered. Note that the direct Urca density threshold for the
  zero viscosity case lies outside of the shown density range.}
\label{fig:micro}
\end{figure}

In order to be consistent with the original fully temperature and
composition dependent EOS, $Y_e^{\mathrm{eq}}$ must be chosen accordingly
from the same original Table. However, note that by construction we are
able to treat the electron fraction of $\beta$-equilibrated matter as a
free function of rest-mass density. This freedom allows us to explore
systematically different strengths of bulk viscosity. We present
$Y_e^{\mathrm{eq}}$ and the square of its derivative, \ie
$\left(\mathrm{d}Y_e^{\mathrm{eq}}/\mathrm{d}\rho\right)^2$ for all
viscosity cases considered in this work in Fig.~\ref{fig:micro}. Black
lines denote the ``zero'' viscosity case, where the $Y_e^{\mathrm{eq}}$
is obtained from the original Table and is vanishingly small. Note that
the derivative of $Y_e^{\mathrm{eq}}$ is very low for that case leading
to an evolution of the system which is practically indistinguishable from
a perfect-fluid model at the resolutions employed in this work. The low,
medium and high viscosity cases are shown in green, blue and red,
respectively. They are determined by the piecewise linear function
\begin{align}\label{eq:yebeta}
Y_e^{\mathrm{eq}} \!=\! \left\{\begin{array}{ccc}
Y_e^{\mathrm{sat}}\rho/\rho_{\mathrm{sat}}\,, & \!\rho \leq
\rho_{\mathrm{sat}}\,,\\
\!\!\!\left(\frac{\rho-\rho_{\mathrm{sat}}}{\rho_{\mathrm{end}} - \rho_{\mathrm{sat}}}\right)
\!\left(Y_e^{\mathrm{end}} \! - \!
Y_e^{\mathrm{sat}}\right) \!+\! Y_e^{\mathrm{sat}}\,,
~~\rho_{\mathrm{sat}} \! < \! &\! \rho \! < \! \rho_{\mathrm{end}}\,, \\
\!\!\!Y_e^{\mathrm{end}}\,, & \!\rho \geq \rho_{\mathrm{end}}\,.
\end{array}\right.
\end{align}
The electron fraction at nuclear saturation density
$Y_e^{\mathrm{sat}}\approx 0.04625$ is determined from chiral effective
field theory \cite{Keller2023} and $\rho_{\mathrm{sat}}$ denotes the
nuclear saturation density. The parameters $\rho_{\mathrm{end}}$ and
$Y_e^{\mathrm{end}}$ are used to systematically increase the slope of the
electron fraction in $\beta$-equilibrium and are shown in
Tab.~\ref{tab:params}. Note that the highest density achieved in our
simulations is approximately $\rho\approx4.88\,\rho_{\mathrm{sat}}$ such
that $Y_e\lesssim0.5$ is satisfied for all cases studied in this work.

\begin{table}
\centering
\begin{tabular}{lccccccccc}
& $k_e$ & $Y_e^{\mathrm{sat}}$ & $\rho_{\mathrm{end}}$ &
$Y_e^{\mathrm{end}}$ & $\gamma_0$ & $U$ & $\alpha$ & $T_{\mathrm{free}}$
& $T_{\mathrm{trap}}$  \\
Model
& $\mathrm{[c^2]}$ && $[\rho_{\mathrm{sat}}]$ && $[\mathrm{MeV}]$ & & &
$[\mathrm{MeV}]$ & $[\mathrm{MeV}]$ \\
\hline
zero visc. & $0.6$ & $-$ & $-$  & $-$ & $15$ & $8.21$ & $5$ & $10$ & $15$ \\[3pt]
low visc.  & $0.6$ & $0.046$ & $7$ & $7\,Y_e^{\mathrm{sat}}$ & $15$ &
$2.4$ & $5$ & $10$ & $15$ \\[3pt] 
med. visc. & $0.6$ & $0.046$ & $7$ & $0.5$ & $15$ & $1.86$ & $5$ &
$10$ & $15$ \\[3pt]
high visc. & $0.6$ & $0.046$ & $7$ & $0.75$ & $15$ & $1.55$ & $5$ &
$10$ & $15$ \\
\hline
\end{tabular}
\caption{Simulation parameters.}
\label{tab:params} 
\end{table}

\begin{figure}
\includegraphics[width=0.49\textwidth]{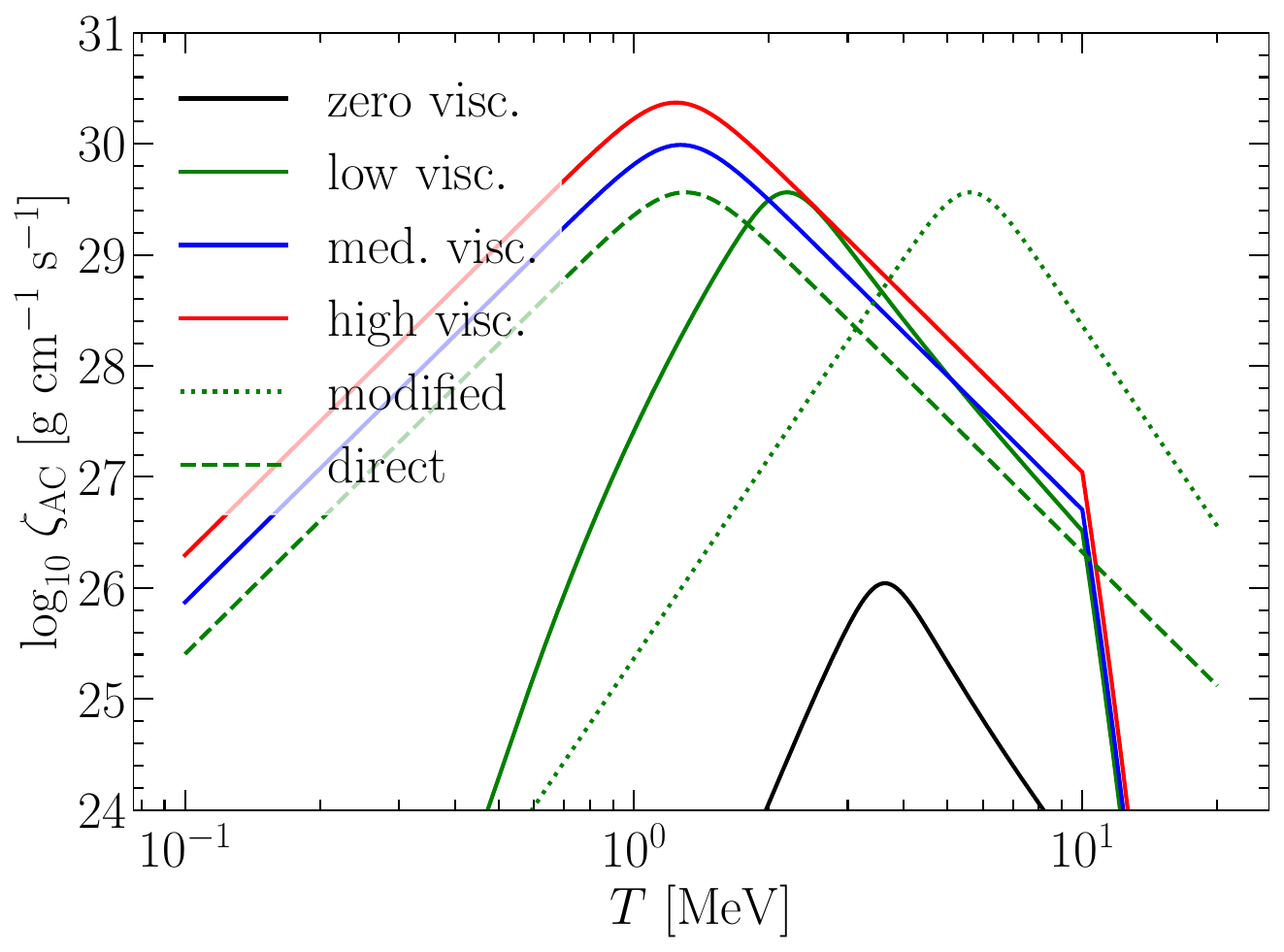}
\caption{Bulk viscosity experienced by perturbative harmonic density
  oscillations of frequency $f = \omega/2\pi = 1~\mathrm{kHz}$ at $\rho =
  2\rho_{\mathrm{sat}}$ for the zero (black solid line), low (green
  solid line), medium (blue solid line) and high viscosity case (red
  solid line). Calculations employing solely direct or modified Urca
  rates are shown for the low viscosity case in dashed and solid green
  lines, respectively.}
\label{fig:aczeta}
\end{figure}

\noindent\emph{Reaction rates and transport coefficients.~}

The linearized \textit{direct} and \textit{modified} Urca reaction rates
are obtained from \cite{Camelio2022_a} in Eqs.~(77) and (79) and are
denoted by $\mathcal{R}^{\mathrm{d}}$ and $\mathcal{R}^{\mathrm{m}}$,
respectively. We now want to combine $\mathcal{R}^{\mathrm{d}}$ and
$\mathcal{R}^{\mathrm{m}}$ in a suitable way by effectively taking into
account the direct Urca threshold density $\rho^{\mathrm{d}}$, \ie the
density below which direct Urca reactions are kinematically forbidden at
zero temperature. Direct Urca processes are Boltzmann suppressed for
$\rho<\rho^{\mathrm{d}}$ which results in factors of
$\exp(-|\gamma_{i}|/T)$ in the rate expressions, where $\gamma_{i}(p)$
defines the single particle free energy of particle $i$, see
\cite{Alford2018a}.
%
%

We model the direct Urca density threshold effectively by taking into
account the Boltzmann suppression for a parametrized single particle free
energy $\gamma$ such that the total rate can be written as
\begin{align}
\mathcal{R}^{\mathrm{tot}}&=\mathcal{R}^{\mathrm{m}}+\exp(-\gamma/T)\mathcal{R}^{\mathrm{d}}\,,\\
\gamma &=
\mathrm{max}\left[0,-\gamma_0U^{-2}\left(\rho^2/\rho^{2}_{\mathrm{sat}}-U^2\right)\right]\,,
\end{align}
where the two constant parameters $\gamma_0$ and $U$ denote the effective
single particle free energy for $\rho \ll \rho^{\mathrm{d}}$ ($\gamma_0
\sim 15-25~\mathrm{MeV}$) and the direct Urca density threshold
$U:=\rho^{\mathrm{d}}/\rho_{\mathrm{sat}}$ expressed in terms of the
nuclear saturation density.
 
Furthermore, recent studies \cite{Espino2024b} have established that at
the high temperatures reached in BNS mergers, \eg $T \gtrsim
10~\mathrm{MeV}$, neutrinos become effectively \textit{trapped} inside
the merger remnant and can particpate in Urca reactions as reactants. As
shown in \cite{Alford2019} neutrino trapping can dramatically increase
the direct Urca reaction rates by several orders of magnitude. We model
the effect of neutrino trapping on the reaction rate effectively by
increasing the total reaction rate by a constant factor of $10^{\alpha}$
over a defined temperature interval
$\left(T_{\mathrm{free}},T_{\mathrm{trap}}\right)$, where we use the
neutrino transparent rate $\mathcal{R}^{\mathrm{tot}}$ for
$T<T_{\mathrm{free}}$, the increased rate
$\mathcal{R}^{\mathrm{tot}}\times10^{\alpha}$ for $T>T_{\mathrm{trap}}$
and an exponential interpolation in the transition region.

\begin{figure}[t!]
\includegraphics[width=0.5\textwidth]{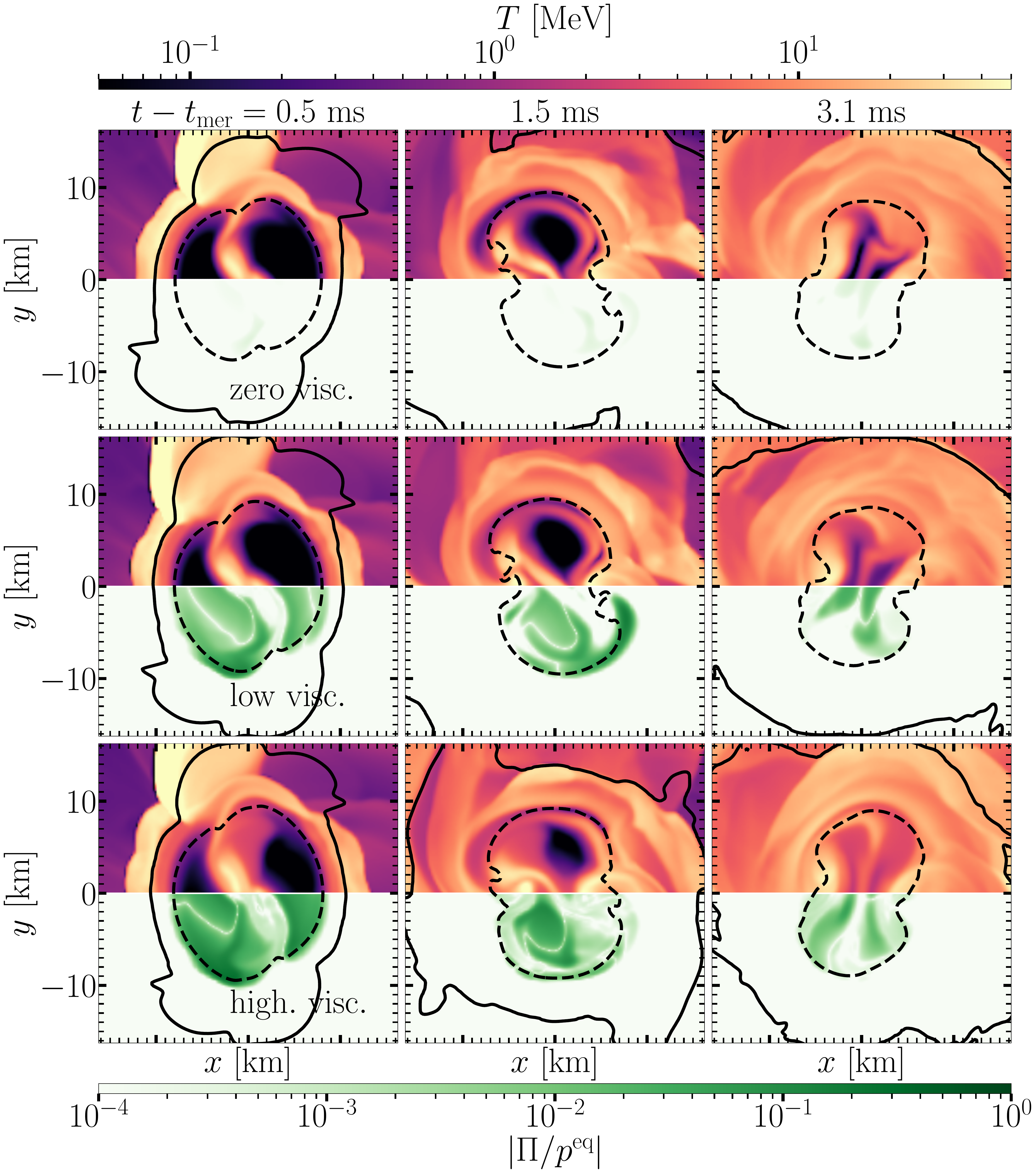}
\caption{\textit{Top half panels:} Temperature distribution in the
  $(x,y)$ plane at $t-t_{\mathrm{mer}}=0.5$ (first column), $1.5$ (second
  column) and $3.1$ (third column). Displayed are the zero, low and high
  viscosity cases in the first, second and third row,
  respectively.  \textit{Bottom half panels:} Same as top half panels but
  for the absolute value of the ratio between the bulk viscous pressure
  and the equilibrium pressure $|\Pi/p^{\mathrm{eq}}|$. Solid and dashed
  lines represent density contour lines at $\rho = 5 \times
  10^{-3} \, \rho_{\mathrm{sat}}$ and $1.82\,\rho_{\mathrm{sat}}$,
  respectively. Note that the transport coefficients transition to their
  atmopshere values in this density range, \ie $\zeta\approx 0$ for
  $\rho\leq 5\times10^{-3}\,\rho_{\mathrm{sat}}$. Hence, the non-perfect
  fluid transitions smoothly to a perfect fluid for densities below
  $\rho=1.82\,\rho_{\mathrm{sat}}$.}
\label{fig:snapshots}
\end{figure}

Finally, following again \cite{Gavassino2021,Camelio2022_a,
  Camelio2022_b}, the bulk viscosity, \ie $\zeta = {\rho}^4
m_{\mathrm{b}}^{-2} \Xi^{-1} \left(\partial Y_e^{\mathrm{eq}}/\partial
\rho \right)_s^2$, and the relaxation time, \ie $\tau = \rho
m_{\mathrm{b}}^{-2}\Xi^{-1} \left(\partial^2\epsilon/\partial
Y_e^{\phantom{e}2} \right)_{\rho,s}^{-1}$, for our system of $npe$-matter
can be computed. Here, we define $\Xi = \left.\left(\partial
\mathcal{R}/\partial \mathbb{A}\right)_{\rho,s}\right|_{\mathbb{A}=0}$.
We observe that the magnitude of $(\zeta/\tau)$ can be increased by
increasing the \textit{slope} of $Y_e^{\mathrm{eq}}$. This has important
implications as $(\zeta/\tau)$ is independent of the reaction rate
$\mathcal{R}$. Hence, we compute the bulk viscosity experienced by
perturbative harmonic density oscillations \cite{Yang2023}, \ie $
\zeta_{_{\mathrm{AC}}}(\omega):=\zeta(1+\omega^2\tau^2)^{-1} =
(\zeta/\tau)\left[(\omega\tau)^{-1}+\omega\tau\right]^{-1}\omega^{-1}\,,$
for a given angular frequency $\omega$ and rest-mass density. We show
results, which are representative for the postmerger phase of a BNS
system, in Fig.~\ref{fig:aczeta} where we choose $\omega = 2\pi f$ with
$f=1~\mathrm{kHz}$ and $\rho=2\,\rho_{\mathrm{sat}}$.

We observe that all curves shown in Fig.~\ref{fig:aczeta} display a
common resonant form where the peak is located at a temperature when
$\omega\tau=1$. Second, as anticipated, we observe a systematic increase
of $\zeta_{_{\mathrm{AC}}}$ for the different viscosity cases used in
this work. This can be easily understood by considering the scaling of
$(\zeta/\tau)$ with the derivative of $Y_e^{\mathrm{eq}}$ and its
independence of $\mathcal{R}$, which is the only temperature dependent
term in $\zeta$ and $\tau$, as well as the fact that $\tau=\omega^{-1}$
at the resonant peak. Third, the impact of employing only
$\mathcal{R}^{\mathrm{d}}$ or $\mathcal{R}^{\mathrm{m}}$ in the
calculations for the low viscosity case show that direct (modified) Urca
reactions decrease (increase) the temperature of the resonant maximum
while keeping its magnitude constant. Fourth, for the medium and high
viscosity case the direct Urca density threshold falls below
$\rho^{\mathrm{d}}<2\,\rho_{\mathrm{sat}}$, see Fig.~\ref{fig:micro}. As
a result, $\zeta_{_{\mathrm{AC}}}$ is dominated by
$\mathcal{R}^{\mathrm{d}}$ for these cases. Finally, we observe a strong
decrease in $\zeta_{_{\mathrm{AC}}}$ for $T>10~\mathrm{MeV}$ as the
result of neutrino trapping.

\noindent\emph{Viscosity distribution.~}
We next discuss the evolution of the bulk-viscous pressure
$\Pi$ and its close correlation with the temperature of the fluid in
Fig.~\ref{fig:snapshots}. In the top half panels of
Fig.~\ref{fig:snapshots}, we show the temperature distribution in the
$(x,y)$ plane at representative times (columns) for the zero, low and
high viscosity case (rows). In the bottom half panels of
Fig.~\ref{fig:snapshots} instead, we show the absolute value of the ratio
between the bulk viscous pressure and the equilibrium pressure
$|\Pi/p^{\mathrm{eq}}|$. Overall, both quantities have a strong
anisotropic distribution and we observe a strong correlation between low
temperatures and large $|\Pi/p|$. This is not surprising as at large
temperatures neutrinos become trapped and effectively increase the
reaction rate of direct Urca processes thereby decreasing the bulk
viscosity.

\end{document}